\documentclass[a4paper,12pt]{article}
\usepackage[dvips]{graphicx}

\usepackage{amsmath}

\newcommand{\D}{\ensuremath{\mathrm{d}}}

\newcommand{\Di}[1]{\ensuremath{\!\!\mathrm{d}#1\,}}

\newcommand{\thickone}{\mbox{$1\!\!1$}}

\newcommand{\halfint}[1]{\ensuremath{\genfrac{\lfloor}{\rfloor}{}{}{#1}{2}}}

\newcommand{\qbinom}[2]{\ensuremath{\genfrac{[}{]}{0pt}{}{\,#1\,}{\,#2\,}_{\!q}}}

\newcommand{\bra}[1]{\ensuremath{\langle #1 |}}

\newcommand{\ket}[1]{\ensuremath{| #1 \rangle}}

\newcommand{\braket}[2]{\ensuremath{\langle #1 | #2 \rangle}}

\newcommand{\braopket}[3]{\ensuremath{\langle #1 | #2 | #3 \rangle}}

\newcommand{\basis}[1]{\ensuremath{\{\ket{#1}\}}}

\newcommand{\invq}{\ensuremath{q^{-1}}}

\begin{document}


\begin{center}
  \quad\\
  {\LARGE Exact solution of a partially asymmetric exclusion model
    using a deformed oscillator algebra} \vskip1cm {\large
    R.~A.~Blythe{\footnote {\tt r.a.blythe@ed.ac.uk}},
    M.~R.~Evans{\footnote {\tt m.r.evans@ed.ac.uk}},
    F.~Colaiori{\footnote {{\tt fran@a13.ph.man.ac.uk}\quad
        Present address: Dept.\ of Physics and Astronomy, University of
        Manchester, Oxford Road, Manchester M13 9PL, U.K.}}  and
    F.~H.~L.~Essler{\footnote {\tt fabs@hubbard.ph.kcl.ac.uk}}}
\end{center}
\vskip1cm
\begin{center}
$^{1,}{^2}$\, Department of Physics and Astronomy, University of Edinburgh,\\
       Mayfield Road, Edinburgh EH9 3JZ, U.K. \\[2ex]
$^3$\, Theoretical Physics, University of Oxford,\\
       1 Keble Road,  Oxford OX1 3NP, U.K.\\[2ex]
$^4$\, Department of Physics, King's College London,\\
       Strand, London WC2R 2LS, U.K.\\
\vskip1cm
October 15th, 1999\\
Revised December 15th, 1999
\end{center}

\vfill


\begin{abstract}
  We study the partially asymmetric exclusion process with open
  boundaries. We generalise the matrix approach previously used to
  solve the special case of total asymmetry and derive exact
  expressions for the partition sum and currents valid for all values
  of the asymmetry parameter $q$. Due to the relationship between the
  matrix algebra and the $q$-deformed quantum harmonic oscillator
  algebra we find that $q$-Hermite polynomials, along with their
  orthogonality properties and generating functions, are of great
  utility. We employ two distinct sets of $q$-Hermite polynomials, one
  for $q<1$ and the other for $q>1$. It turns out that these
  correspond to two distinct regimes: the previously studied case of
  forward bias ($q<1$) and the regime of reverse bias ($q>1$) where
  the boundaries support a current opposite in direction to the bulk
  bias.  For the forward bias case we confirm the previously proposed
  phase diagram whereas the case of reverse bias produces a new phase
  in which the current decreases exponentially with system size.
\end{abstract}

\vfill

\newpage


\section{Introduction}

The asymmetric simple exclusion process (ASEP) \cite{Liggett} is a
much studied model both from a mathematical and physical viewpoint.
The model comprises particles hopping in a preferred direction on a
lattice with hard core exclusion imposed. In the mathematical
literature the interest lies in its being a simple realisation of
interacting Markov processes \cite{Spitzer} and much progress has been
made in proving existence theorems, invariant measures and
hydrodynamic limits \cite{Liggett}. Early applications concerned
biophysical problems such as single-filing constraint in transport
across membranes \cite{Heckmann} and the kinetics of biopolymerisation
\cite{MGP}.  More recently the ASEP has achieved the status of a
fundamental non-equilibrium model due to its intimate relation to
growth phenomena and the KPZ equation \cite{Krug97}, the problem of
directed polymers in a random media \cite{HHZ} and its use as a
microscopic model for driven diffusive systems \cite{SZ} and shock
formation \cite{JLbook}.  Last but not least many traffic flow models
are based on variants of the ASEP \cite{SW}. Adding to its appeal is
the fact that many exact results have been obtained, particularly in
one dimension, allowing an analytical understanding of the
non-equilibrium phenomena exhibited \cite{DerridaReview}.

In recent years the open boundary problem, where particles attempt to
enter at the left of a one-dimensional lattice of $N$ sites with rate
$\alpha$, hop to the right under a hard core exclusion constraint and
exit at the right with rate $\beta$, has been of considerable
interest. It was first pointed out by Krug that boundary-induced phase
transitions can occur \cite{Krug} and it has further been shown that
spontaneous symmetry breaking takes place when two oppositely moving
particle species are introduced \cite{EFGMPRL}.

The door was opened to the analytical study of the open boundary model
and its non-equilibrium steady state in \cite{DDM}. There, exact
recursion relations on the steady state weights were obtained and
density profiles and currents were worked out exactly for the case
$\alpha=\beta=1$. Also a mean field phase diagram in the
$\alpha$--$\beta$ plane was derived.  Then the recursion relations
were used to calculate correlation functions for the case $\alpha=
\beta=1$ \cite{DE93}. In \cite{DEHP} a different method, to be
referred to as the matrix approach, was proposed: it was shown that
the steady state weights can written as a product of matrices which
are in general of infinite dimension and non-commuting. The matrices
obey algebraic rules which replace the recursion relations found in
\cite{DDM}. This approach gives the full solution of the model
including the phase diagram and density profiles and allows, in
principle, calculation of all equal time correlation functions.
Further it admits generalisation to other models.  It should be noted
that the phase diagram and density profiles were also obtained
independently working from the recursion relations \cite{SD}.
Subsequently the matrix approach has been used to solve new models
with several species \cite{DJLS,EFGM,AHR,Karimipour} and various
updating schemes \cite{RS,ERS,dGN}, to calculate shock profiles on an
infinite system \cite{DLS} and to recover some previously known
results \cite{HS}.

The phase diagram obtained for the one species open boundary case
\cite{DDM,DEHP,SD} comprises three phases: a low density phase, a high
density phase and a maximal current phase where generic long range
correlations occur.  This phase diagram appears quite robust for
driven diffusive systems with open boundaries \cite{KSKS}. Also it has
been shown that different types of updates the same generic phase
diagram pertains, except in simple cases of deterministic bulk
dynamics where the maximal current phase is absent \cite{Schutz93,TE}.

The partially asymmetric exclusion process is a generalisation of the
model where particles are able to hop to the left as well as to the
right.  The case of partial asymmetry is of interest since it allows
one to interpolate between symmetric exclusion which can have
equilibrium steady states and the far from equilibrium asymmetric
system.  In the context of growth phenomena the two different systems
are described by Edwards-Wilkinson and KPZ universality classes
\cite{Krug} and the crossover phenomena has been of interest
\cite{Kim,DM}.

A parameter $q$ gives the ratio of hopping rates to the left and to
the right; thus $q=0$ recovers the totally asymmetric process and
$q=1$ gives the fully symmetric case. Furthermore in the range $0<q<1$
there exists what we shall refer to as the {\em forward bias} regime
where particles hop preferentially to the right and the boundary
conditions are such that a steady state current of particles to the
right is supported. In contrast, when $q>1$ particles can only enter
at the left and leave at the right but hop preferentially to the left
in the bulk. A steady state current of particles to the right can then
be supported by the boundary conditions {\em against} the bulk bias.
This gives rise to a new phase which we analyse for the first time
here. We show that the current decreases exponentially with the length
of the system.  This phase is of interest in the context of `backbend
dynamics' \cite{RB,TB} where, for example, a fluid in a permeable
medium has to traverse a pore oriented against the direction of
gravity.  It also appears that such a reverse bias phase is relevant
to recently reported one-dimensional phase separation \cite{EKKM,AHR}.

In \cite{DEHP} the generalisation of the matrix approach to the
partially asymmetric exclusion process was pointed out. This was used
in \cite{Sandow} to obtain approximate expressions for the current in
the forward bias case for large system size. It was also pointed out
that the matrices for the partially asymmetric case are closely
related to creation and annihilation operators of the $q$-deformed
harmonic oscillator.  In \cite{ER} the quadratic algebra was studied
and curves in the parameter space were deduced for which finite
matrices could be used. Along these curves exact expressions for
physical quantities, such as the current and correlation length, can
be computed easily.  Also, in the symmetric case $q=1$ the steady
state is straightforward to solve \cite{Spohn,SMW,SS}. However exact
expressions for all system sizes and parameters have remained elusive.

In this work we build on the relationship of the quadratic algebra
with the $q$-deformed harmonic oscillator algebra to calculate exact
expressions for the current for all system sizes and parameter values.
The basic approach is to use known properties, such as orthogonality
and generating functions, of the $q$-deformed Hermite polynomials
associated with the quadratic algebra.

This method was employed independently by \cite{Sasamoto1} to study
the case of forward bias ($q<1$) and a phase diagram for this regime
was obtained by examining the asymptotics of the normalisation
(partition sum). We have been able to consider for the first time the
case $q>1$ for which a less well-known set of $q$-Hermite polynomials
is required \cite{Askey}. Furthermore we obtain as a main new result
of our work an exact, explicit expression for the normalisation valid
for \textit{all} $q$ which encompasses all regimes: $q=0$ (total
asymmetry), $q<1$ (forward bias), $q=1$ (symmetric) and $q>1$ (reverse
bias).  This exact expression allows all physical quantities of the
model to be evaluated for any system size and we use it here to obtain
the asymptotic form of the current in the reverse bias regime.

The paper is organised as follows: in section~2 we define the model we
consider and in section~3 we review the matrix approach and the
related quadratic algebra. In section~4 we discuss the $q$-deformed
harmonic oscillator and its relation to the quadratic algebra of the
present problem.  In particular, we present relevant facts such as the
generating functions and orthogonality relations for the $q$-Hermite
polynomials.  In section 5 we derive our main results, which are exact
expressions for the normalisation (partition sum).  In
section~\ref{sec:intrep:q<1} we derive an integral expression valid
for $q<1$ (\ref{eqn:Zint:q<1:short}) and in
section~\ref{sec:intrep:q>1} we derive an integral expression valid
for $q>1$ (\ref{eqn:Zint:q>1:short}). We then obtain in
section~\ref{sec:sumformula} a finite sum expression valid for all $q$
(\ref{eqn:Zsumrep}).  In section~\ref{sec:phasediag} we use the exact
expressions to calculate the asymptotic behaviour of the current and
the phase diagram.  We conclude in section~\ref{sec:conclude} with a
discussion.

\section{Model definition}

The microscopic dynamics of the model are specified by four rates at
which certain events can occur. For a rate $\lambda$ associated with a
particular event, the probability that the event happens in an infinitesimal
time interval $\D t$ is $\lambda \D t$. Furthermore, moves that would
lead to two particles simultaneously occupying a single lattice site
are prohibited due to the hard-core repulsion between them.

The events defined in the model and the rates at which they take place
are as follows.

\begin{tabular}{lc}
\textbf{Event}                                 & \textbf{Rate} \\
Particle inserted onto the left boundary site  & $\alpha$      \\
Particle removed from the right boundary site  & $\beta$       \\
Particle hops by one site to the right         & $1$           \\
Particle hops by one site to the left          & $q$
\end{tabular}

\begin{figure}[htb]
\begin{center}
\includegraphics[scale=0.9]{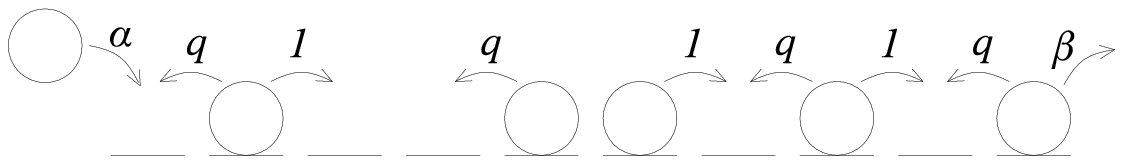}
\end{center}
\caption{\label{fig:PASEPrules}Typical particle configurations and
  allowed moves in the model.}
\end{figure}

Figure \ref{fig:PASEPrules} shows two typical particle
configurations on a small lattice along with the allowed moves
and their rates.

As only three of the rates are independent, we have set the right hopping
rate to $1$ with no loss of generality in the following analysis.

Later, in section~\ref{sec:conclude}, we will consider a more general
parameter space where particles can also enter at the right and exit
at the left.

\section{The matrix product formulation and its quadratic algebra}
\label{sec:DEalgebra}

In this section we review the matrix approach to finding the steady
state of the model. We present here the bare essentials of the method
and refer the reader elsewhere for more detailed descriptions of the
technique \cite{DEHP,DerridaReview,Krebs}.

Consider first a configuration of particles $\mathcal{C}$ and its
steady-state probability $P(\mathcal{C})$. We use as an ansatz for
$P(\mathcal{C})$ an ordered product of matrices $X_1 X_2 \ldots X_N$
where $X_i=D$ if site $i$ is occupied and $X_i=E$ if it is empty. To
obtain a probability (a scalar value) from this matrix product, we
employ two vectors $\bra{W}$ and $\ket{V}$ in the following way:
\begin{equation}
\label{eqn:Pansatz}
P(\mathcal{C})=\frac{\braopket{W}{X_1 X_2 \ldots X_N}{V}}{Z_N} \;.
\end{equation}
The factor $Z_N$ is included to ensure that $P(\mathcal{C})$ is
properly normalised. This latter quantity, analogous to a partition
function, has the following simple matrix expression through which a
new matrix $C$ is defined:
\begin{equation}
\label{eqn:Zdef}
Z_N=\braopket{W}{(D+E)^N}{V}=\braopket{W}{C^N}{V} \;.
\end{equation}

Note that if $D$ and $E$ do not commute $P(\mathcal{C})$ is a
function of both the number and position of particles on the lattice,
as expected for a non-trivial steady state. The algebraic properties
of the matrices can be deduced from the Master equation for the
process \cite{DEHP}. It can be shown that sufficient conditions for 
equation (\ref{eqn:Pansatz}) to hold are
\begin{eqnarray}
\label{eqn:DEcommute}
DE-qED &=& D+E \\
\label{eqn:EonW}
\alpha \bra{W}E &=& \bra{W} \\
\label{eqn:DonV}
\beta D \ket{V} &=& \ket{V} \;.
\end{eqnarray}

One can also write expressions for ensemble-averaged quantities in
terms of matrix products. For example the current of particles $J$
through the bond between sites $i$ and $i+1$ is given by
\label{eqn:jdef}
\begin{equation}
J= \frac{\braopket{W}{C^{i-1}(DE-qED)C^{N-i-1}}{V}}{Z_N}
= \frac{Z_{N-1}}{Z_N}
\end{equation}
where the last equality follows from relation (\ref{eqn:DEcommute}).
We see that, as expected in the steady state, the current is
independent of the bond chosen. Also, the mean occupation number
(density) of site $i$ may be written as
\begin{equation}
\label{eqn:taudef}
\tau_i = \frac{\braopket{W}{C^{i-1}DC^{N-i}}{V}}{Z_N} \;.
\end{equation}
Our task now is to evaluate the matrix products in the above
expressions for $Z_N$, $J$ and $\tau_i$ by applying the rules
(\ref{eqn:DEcommute}--\ref{eqn:DonV}).

In \cite{DEHP} the case $q=0$ was treated by using
(\ref{eqn:DEcommute}) repeatedly to `normal-order' matrix products:
that is, to obtain an equivalent sum of products in which all $E$
matrices appear to the left of any $D$ matrices.  Then finding a
scalar value from (\ref{eqn:EonW}) and (\ref{eqn:DonV}) is
straightforward. For example one can develop (\ref{eqn:Zdef}) as
\begin{equation}
\label{eqn:canonicalform}
Z_N = \braopket{W}{C^N}{V}=
\sum_{n,m}a_{N,n,m} \braopket{W}{E^nD^m}{V} = \braket{W}{V}
\sum_{n,m}a_{N,n,m} \alpha^{-n} \beta^{-m}
\;.
\end{equation}
The difficulty with this approach lies in the combinatorial problem of
finding the coefficients $a_{N,n,m}$. An alternative approach proposed
in \cite{DEHP} is to find an explicit representation of $C$ and
decompose the vectors $\bra{W}$ and $\ket{V}$ onto the eigenbasis of
$C$ to evaluate the normalisation.

In the present work we employ mainly the latter approach to derive an
expression for the normalisation which is valid over a restricted
range of the model parameters. We will later compare this with the
canonical form (\ref{eqn:canonicalform}) to find an expression which
does not depend on the chosen representation and is therefore
generally valid. The representation of $C$ that we use is intimately
related to the $q$-oscillator algebra and the eigenbasis of $C$ is
constructed from $q$-Hermite polynomials.

\section{The $q$-deformed harmonic oscillator and its relevance}
\label{sec:qalgebra}

Much progress can be made if the algebra of the previous section is
written in terms of that of $q$-deformed quantum harmonic oscillator
\cite{Macfarlane}. The relationship central to this algebra is the
$q$-deformed commutator
\begin{equation}
\hat{a}\hat{a}^\dagger-q\hat{a}^\dagger\hat{a}=1
\label{eqn:qcommute}
\end{equation}
where the operators $\hat{a}$ and $\hat{a}^\dagger$ operate on basis
vectors $\ket{n}$ (with $n=0,1,2,\ldots$) as follows:
\begin{eqnarray}
\label{eqn:adaggeronn}
\hat{a}^\dagger \ket{n} &=& \left( \frac{1-q^{n+1}}{1-q}
\right)^{\!\frac{1}{2}} \ket{n+1} \\
\label{eqn:aonn}
\hat{a} \ket{n} &=& \left( \frac{1-q^n}{1-q}
\right)^{\!\frac{1}{2}} \ket{n-1} \\
\hat{a} \ket{0} &=& 0 \;.
\end{eqnarray}

In terms of these new operators, the matrices introduced in the
previous section can be written as
\begin{eqnarray}
\label{eqn:Dofa}
D &=& \frac{1}{1-q}+\frac{1}{\sqrt{1-q}}\:\hat{a} \\
\label{eqn:Eofa}
E &=& \frac{1}{1-q}+\frac{1}{\sqrt{1-q}}\:\hat{a}^\dagger
\end{eqnarray}
and one finds using (\ref{eqn:qcommute}) that (\ref{eqn:DEcommute}) is
satisfied.

We now have an explicit representation of the original $D$ and $E$
matrices in the oscillator's ``energy'' eigenbasis \basis{n}. Thus one
may combine equations (\ref{eqn:Dofa}), (\ref{eqn:aonn}) and
(\ref{eqn:DonV}) to find the corresponding representation of the
vector $\ket{V}$:
\begin{equation}
\label{eqn:Vofn:long}
\braket{n}{V}=\frac{v^n}{\prod_{j=1}^{n} \sqrt{(1-q^j)}}
\end{equation}
where we have set $\braket{0}{V}=1$ and $v$ is the following
combination of the model parameters:
\begin{equation}
\label{eqn:vofbeta}
v=\frac{1-q}{\beta}-1
\end{equation}

The denominator of (\ref{eqn:Vofn:long}) is more conveniently written
in terms of a `$q$-shifted factorial'. This is defined through
\begin{eqnarray}
\label{eqn:qfacdef}
(a;q)_n &=& \prod_{j=0}^{n-1} (1-aq^j) \\
(a;q)_0 &=& 1 \;.
\end{eqnarray}
We will later encounter products of these factorials for which we
shall use the shorthand employed by \cite{GaspRah}:
\begin{equation}
(a,b,c,\ldots;q)_n=(a;q)_n (b;q)_n (c;q)_n \ldots \;.
\end{equation}

We can now use this compact notation to write expressions for both
$\braket{n}{V}$ and $\braket{W}{n}$, the route to the latter being the
same as that to find $\braket{n}{V}$:
\begin{eqnarray}
\label{eqn:Vofn}
\braket{n}{V}=\frac{v^n}{\sqrt{(q;q)_n}} \\
\label{eqn:Wofn}
\braket{W}{n}=\frac{w^n}{\sqrt{(q;q)_n}}
\end{eqnarray}
where $v$ is given by (\ref{eqn:vofbeta}) and
\begin{equation}
\label{eqn:wofalpha}
w=\frac{1-q}{\alpha}-1 \;.
\end{equation}

We see that the representation of $D$ and $E$ (\ref{eqn:Dofa}) and
(\ref{eqn:Eofa}) breaks down for certain choices of the model
parameters. Firstly, for $q=1$ (symmetric exclusion) a number of
singularities appear.  Secondly, if $v>1$ and $q<1$, the vector
element $\braket{n}{V}$ (\ref{eqn:Vofn}) is unbounded from above as $n
\to \infty$; similarly with $\braket{W}{n}$ (\ref{eqn:Wofn}) when
$w>1$ and $q<1$. We consider for the moment only those regions of
parameter space where this representation converges, and discuss the
generalisation to the remaining areas in section~\ref{sec:sumformula}.

We persevere with the relationship between the original quadratic
algebra and the $q$-oscillator algebra for the reason we now explain.
We introduced earlier a matrix $C$ which appears in the expressions
for the mean particle density and current. We now see that this matrix
can be written as a linear combination of the identity $\thickone$ and
the ``co-ordinate'' operator $\hat{x}=\hat{a}+\hat{a}^\dagger$:
\begin{equation}
\label{eqn:Cofx}
C = D+E = \frac{2}{1-q}\:\thickone + \frac{1}{\sqrt{1-q}}\:\hat{x}
\;.
\end{equation}

The eigenstates of the oscillator in the co-ordinate representation
are known --- in analogy with the solutions of the undeformed
oscillator they are called the continuous $q$-Hermite polynomials
\cite{GaspRah}. Clearly, the eigenvectors of $C$ are the
same as those for $\hat{x}$ and therefore knowledge of them permits
diagonalisation of $C$. As this is a major step towards obtaining the
exact solution of the model, it is worth spending a little time
discussing the $q$-Hermite polynomials.

The recursion relation for the polynomials follows after a suitable
definition of the operator $\hat{x}$ on its eigenbasis $\basis{x}$:
\begin{equation}
\label{eqn:xonx}
\hat{x}\ket{x}=\frac{2x}{\sqrt{1-q}}\ket{x} \;.
\end{equation}
From this and equations (\ref{eqn:aonn}) and (\ref{eqn:adaggeronn}) we
find
\begin{equation}
\label{eqn:hprecrel}
2x\braket{x}{n} = \sqrt{1-q^n}\: \braket{x}{n-1} + \sqrt{1-q^{n+1}}\:
\braket{x}{n+1} \; .
\end{equation}
Explicit formul\ae\ for $\braket{x}{n}$ can be found using a
generating function technique, the details of which differ slightly
depending on whether $q<1$ or $q>1$. Here we present the results which
will be most useful later; derivations are given in appendix A.

When $q<1$ we make a change of variable
\begin{equation}
\label{eqn:xoftheta}
x=\cos\theta \;.
\end{equation}
The $q$-Hermite polynomials can now be defined as
$\braket{\theta}{n}$, that is, the projection of the oscillator energy
eigenstate $\ket{n}$ onto the position basis $\bra{\theta}$. The
generating function
\begin{equation}
\label{eqn:genfunc:q<1}
G(\theta,\lambda)=\sum_{n=0}^{\infty} \frac{\lambda^n}{\sqrt{(q;q)_n}}
\braket{\theta}{n}
\end{equation}
can be expressed as an infinite product
\begin{equation}
G(\theta,\lambda)=\frac{1}{(\lambda e^{i\theta}, \lambda
  e^{-i\theta};q)_\infty}
\end{equation}
when $|\lambda|<1$. An explicit form of the $q$-Hermite polynomial
$\braket{\theta}{n}$ can be determined from the generating function
and is presented in appendix A, equation (\ref{eqn:HP:q<1}). It can
also be shown \cite{GaspRah} that the set of $q$-Hermite polynomials
are orthogonal with respect to a weight function $\nu(\theta)$. That
is
\begin{equation}
\label{eqn:orthog:q<1}
\int_0^\pi \Di{\theta} \braket{n}{\theta} \nu(\theta) \braket{\theta}{m}
=\delta_{n,m}
\end{equation}
where
\begin{equation}
\nu(\theta)=\frac{(q,e^{2i\theta},e^{-2i\theta};q)_\infty}{2\pi}   \;.
\end{equation}

Similar results emerge when $q>1$ under a different change of
variable
\begin{equation}
x=i\sinh u
\end{equation}
with a suitably re-defined generating function valid for all $\lambda$:
\begin{equation}
\label{eqn:genfunc:q>1}
G(u,\lambda)=\sum_{n=0}^{\infty} \frac{\lambda^n}{\sqrt{(q;q)_n}}
\braket{u}{n} = (i\invq\lambda e^u,-i\invq\lambda e^{-u};\invq)_\infty
\;.
\end{equation}
An explicit expression for $\braket{u}{n}$ is given in
(\ref{eqn:HP:q>1}).  Again a weight function $\nu(u)$ that
orthogonalises the polynomials can be found \cite{Askey}. We write
\begin{equation}
  \int_{-\infty}^{\infty} \Di{u} \braket{n}{u} \nu(u)
  \braket{u}{m}=\delta_{n,m}
\end{equation}
where now
\begin{equation}
\label{eqn:nu:q>1}
\nu(u)=\frac{1}{\ln q}\,\frac{1}{(\invq,-\invq e^{2u},-\invq
  e^{-2u};\invq)_\infty} \;.
\end{equation}

\section{Exact expressions for the normalisation $Z_N$}

\subsection{Integral representation for $q<1$}
\label{sec:intrep:q<1}

The relationships in the previous section allow us to obtain integral
representations of matrix products. Here we illustrate how to apply
the procedure to obtain an expression for the normalisation $Z_N$ when
$q<1$. The procedure used to obtain this particular result is also
described by \cite{Sasamoto1}; in the next subsection we extend the
method to the case $q>1$.

First we take the orthogonality relation for the $q$-Hermite
polynomials (\ref{eqn:orthog:q<1}) and use it to form a representation
of the identity matrix:
\begin{equation}
\label{eqn:thickone}
\int_{0}^{\pi} \Di{\theta} \ket{\theta} \nu(\theta) \bra{\theta} =
\thickone \;.
\end{equation}
We now insert this into the expression for the normalisation
(\ref{eqn:Zdef}):
\begin{equation}
Z_N=\int_0^\pi \Di{\theta} \nu(\theta)
\braopket{W}{C^N}{\theta}\braket{\theta}{V} \;.
\end{equation}
By design, the matrix $C$ is acting on its eigenvectors, so using
(\ref{eqn:Cofx}) and (\ref{eqn:xoftheta}) we obtain
\begin{equation}
\label{eqn:Zint}
Z_N=\int_{0}^{\pi} \Di{\theta} \nu(\theta) \braket{W}{\theta} \left(
  \frac{2(\cos\theta+1)}{1-q} \right)^{\!\!N} \braket{\theta}{V} \;.
\end{equation}

It is necessary to decompose the boundary vectors $\bra{W}$ and
$\ket{V}$ onto the $\basis{\theta}$ basis. By inserting a complete set
of the basis vectors $\basis{n}$ we find
\begin{equation}
\braket{\theta}{V}=\sum_{n=0}^{\infty} \braket{\theta}{n}
\braket{n}{V} = \sum_{n=0}^{\infty} \frac{v^n}{\sqrt{(q;q)_n}}
\braket{\theta}{n} \;.
\end{equation}
The final sum in this equation is just the generating function of the
$q$-Hermite polynomials (\ref{eqn:genfunc:q<1}). Thus, when $|v|<1$, we
may write
\begin{equation}
\braket{\theta}{V}=G(\theta,v)=\frac{1}{(ve^{i\theta},ve^{-i\theta};q)_\infty}
\;.
\end{equation}
Similarly, when $|w|<1$ we find
\begin{equation}
\braket{W}{\theta}=G(\theta,w)=\frac{1}{(we^{i\theta},we^{-i\theta};q)_\infty}
\;.
\end{equation}

Putting all this together, we arrive at an exact integral form for the
normalisation
\begin{equation}
\label{eqn:Zint:q<1:short}
Z_N = \left( \frac{1}{1-q} \right)^{\!\!N}
\int_0^\pi \Di{\theta} \nu(\theta) [2(1+\cos\theta)]^N G(\theta,w)
G(\theta,v) 
\end{equation}
which, written out more fully, reads
\begin{equation}
\label{eqn:Zint:q<1:long}
Z_N = \frac{(q;q)_\infty}{2\pi} \left( \frac{1}{1-q} \right)^{\!\!N}
\int_0^\pi \Di{\theta} [2(1+\cos\theta)]^N
\frac{(e^{2i\theta},e^{-2i\theta};q)_\infty}
{(ve^{i\theta},ve^{-i\theta},we^{i\theta},we^{-i\theta};q)_\infty} \;.
\end{equation}

When $|v|>1$ or $|w|>1$ equation (\ref{eqn:Zint:q<1:short}) is
not well-defined because $G(\theta,\lambda)$ does not converge when
$|\lambda|>1$. Rather than finding a representation of the quadratic
algebra that does not suffer from this problem, one can simply
analytically continue the integral (\ref{eqn:Zint:q<1:long}) 
to obtain $Z_N$ when $|v|$ or $|w|$ takes on a value greater
than one. This procedure is carried out in section \ref{sec:phasediag}.

\subsection{Integral representation for $q>1$}
\label{sec:intrep:q>1}

We now apply the procedure of the previous section to find an integral
representation of the normalisation for the case of $q>1$ which has
not previously been considered. The only difference is that we must
use the $q>1$ forms of the generating function and weight function of
the $q$-Hermite polynomials, namely equations
(\ref{eqn:genfunc:q>1}--\ref{eqn:nu:q>1}). We obtain a similar form
for $Z_N$ to (\ref{eqn:Zint:q<1:short}):
\begin{equation}
\label{eqn:Zint:q>1:short}
Z_N = \left( \frac{1}{1-q} \right)^{\!\!N} 
\int_{-\infty}^{\infty} \Di{u} \nu(u) [2(1+i\sinh u)]^N G(u,w) G(u,v)
\;.
\end{equation}
However the full form is somewhat different from (\ref{eqn:Zint:q<1:long}):
\begin{multline}
  Z_N = \frac{1}{\ln q} \frac{1}{(\invq;\invq)_\infty}
  \left( \frac{2}{1-q} \right)^{\!\!N} \\
  \int_{-\infty}^{\infty} \Di{u} (1+i\sinh u)^N \; \frac{(i\invq
    ve^u,-i\invq ve^{-u},i\invq we^u,-i\invq we^{-u};\invq)_\infty}
  {(-\invq e^{2u},-\invq e^{-2u};\invq)_\infty} \;.
\end{multline}
One should note the range of integration is infinite and that, in
contrast to equation (\ref{eqn:Zint:q<1:long}), it cannot be simply
replaced by a closed contour in the complex plane. It is this feature
which makes this integral unsuited to approximation using the
saddle-point method as we discuss section~\ref{sec:revbias}.

\subsection{Explicit formula}
\label{sec:sumformula}

In this section we derive an alternative expression for $Z_N$ which
takes the form of a finite sum rather than an integral and is valid
for all values of the model parameters. Such an expression is
useful for two main reasons: firstly it allows us to extract the
asymptotic form of the normalisation when $q>1$; secondly, as the sum
contains a finite number of terms, it can be evaluated exactly by
numerical means if one wishes to study finite-sized systems.

We will work from the integral for $q<1$, $|v|<1$ and $|w|<1$
(\ref{eqn:Zint:q<1:short}). The first stage of the calculation is to
state an important identity:
\begin{equation}
\label{eqn:qbetaint}
\int_0^\pi \Di{\theta} \nu(\theta) G(\theta,\lambda) G(\theta,v)
G(\theta,w) = \frac{1}{(vw,\lambda v,\lambda w;q)_\infty} \;.
\end{equation}
We do not prove this here but note that it is in fact a special case
of the Askey-Wilson $q$-beta integral \cite{GaspRah}. \footnote{In its
  most general form the Askey-Wilson $q$-beta integral has four
  parameters, whereas identity (\ref{eqn:qbetaint}) has only three.}

We now expand both sides of (\ref{eqn:qbetaint}) in powers of
$\lambda$. We already know the expansion of the left-hand side because
$G(\theta,\lambda)$ is the generating function of the $q$-Hermite
polynomials. The right-hand side may be treated by using another important
identity valid when $|x|<1$, $q<1$ \cite{GaspRah}:
\begin{equation}
\label{eqn:exp:q<1}
\sum_{n=0}^{\infty} \frac{x^n}{(q;q)_n}=\frac{1}{(x;q)_\infty}
\;.
\end{equation}
We find
\begin{equation}
\label{eqn:qbetaexpand}
\frac{1}{(\lambda v,\lambda w;q)_\infty}=
\sum_{n=0}^{\infty} \frac{\lambda^n}{(q;q)_n} \sum_{k=0}^{n}
\qbinom{n}{k} v^{n-k} w^k
\end{equation}
where we have used the $q$-binomial coefficient which is
\begin{equation}
\label{eqn:qbinomial}
\qbinom{n}{k}= \frac{(q;q)_n}{(q;q)_{n-k} (q;q)_k}
\end{equation}
when $0 \leq k \leq n$ and zero otherwise.  In the limit $q \to 1$
the $q$-binomial coefficient is equal to the conventional version
$\binom{n}{k}$ familiar from combinatorics. Thus the last summation in
(\ref{eqn:qbetaexpand}) may be considered a $q$-deformation of the
binomial expansion of $(v+w)^n$. We give this function the symbol
$B_n(v,w;q)$:
\begin{equation}
\label{eqn:Bdef}
B_n(v,w;q)=\sum_{k=0}^{n} \qbinom{n}{k} v^{n-k} w^k \;.
\end{equation}

If we now compare co-efficients of powers of $\lambda$ on both sides
of (\ref{eqn:qbetaint}) we obtain a key result:
\begin{equation}
\label{eqn:keyresult}
\int_0^\pi \Di{\theta} \nu(\theta)
\braket{\theta}{n} G(\theta,v) G(\theta,w) = \frac{1}{(vw;q)_\infty}
\frac{B_n(v,w;q)}{\sqrt{(q;q)_n}}\;.
\end{equation}
This relationship is important because we may take any sufficiently
well-behaved function $f(\theta)$, re-express it it as a sum of
$q$-Hermite polynomials and use (\ref{eqn:keyresult}) to evaluate the
integral
\begin{equation}
\int_0^\pi \Di{\theta} \nu(\theta) f(\theta) G(\theta,v) G(\theta,w) \;.
\end{equation}
Specifically, we can choose $f(\theta)=[2(1+\cos\theta)]^N$ and solve
equation (\ref{eqn:Zint:q<1:short}) exactly.

Expanding the cosine function in this way involves little more than
routine algebra which is detailed in appendix B. The identity which
emerges is
\begin{equation}
\label{eqn:cosexpand}
[2(1+\cos\theta)]^N=\sum_{n=0}^N R_{N,n}(q) \: \sqrt{(q;q)_n}\:
\braket{\theta}{n}
\end{equation}
with
\begin{equation}
\label{eqn:Rdef}
R_{N,n}(q)=\sum_{k=0}^{\halfint{N-n}} (-1)^k \binom{2N}{N-n-2k}
q^{\binom{k}{2}} \left\{ \qbinom{n+k-1}{k-1} + q^k \qbinom{n+k}{k}
\right\}
\end{equation}
which may be alternatively written as
\begin{equation}
\label{eqn:Rdef2}
R_{N,n}(q)=\sum_{k=0}^{\halfint{N-n}} (-1)^k
\left[ \binom{2N}{N-n-2k} - \binom{2N}{N-n-2k-2} \right]
q^{\binom{k+1}{2}} \qbinom{n+k}{k}
\;.
\end{equation}

We may now insert the expansion (\ref{eqn:cosexpand}) into
(\ref{eqn:Zint:q<1:short}) and integrate using (\ref{eqn:keyresult}):
\begin{equation}
\label{eqn:Zsumrep:q<1}
Z_N=\frac{1}{(vw;q)_\infty} \left( \frac{1}{1-q} \right)^{\!\!N}
\sum_{n=0}^N R_{N,n}(q) B_n(v,w;q) \;.
\end{equation}

This exact formula, valid for $q<1$, $|v|<1$ and $|w|<1$ admits
extension to general $q$, $v$ and $w$. We first note that the infinite
product in the prefactor can be replaced with $\braket{W}{V}$, a fact
which follows from (\ref{eqn:exp:q<1}):
\begin{equation}
\frac{1}{(vw;q)_\infty} = \sum_{n=0}^{\infty} \frac{(vw)^n}{(q;q)_n} =
\braket{W}{V} \;.
\end{equation}
We claim that the resulting expression for the normalisation
\begin{equation}
\label{eqn:Zsumrep}
Z_N = \braket{W}{V} \left( \frac{1}{1-q} \right)^{\!\!N}
\sum_{n=0}^N R_{N,n}(q) B_n(v,w;q)
\end{equation}
where $R_{N,n}(q)$ is given by (\ref{eqn:Rdef}) and $B_n(v,w;q)$ by
(\ref{eqn:Bdef}), holds for \textit{all} choices of the model
parameters.

This is justified by observing that once $v$ and $w$ are written in
terms of $\alpha$ and $\beta$ using (\ref{eqn:vofbeta}) and
(\ref{eqn:wofalpha}) we obtain a power series in $\alpha$ and $\beta$
of the same form as (\ref{eqn:canonicalform}).  As discussed in
section \ref{sec:DEalgebra}, an equation with this structure arises
when one re-orders a matrix product directly using the relation
(\ref{eqn:DEcommute}). To perform this direct manipulation, it is not
necessary to employ a specific representation of the quadratic algebra.
Therefore, although the representation we used to derive
(\ref{eqn:Zsumrep}) breaks down for $q=1$, $q<1, |v|>1$ or $q<1,
|w|>1$, we can now say that had we used one which converges in the
region of interest, we would still have obtained equation
(\ref{eqn:Zsumrep}) for the normalisation.

As a check of this formula, let us consider the case $q=0$. Then
(\ref{eqn:Rdef2}) and (\ref{eqn:Bdef}) become
\begin{eqnarray}
R_{N,n}(0)&=&\binom{2N}{N-n}-\binom{2N}{N-n-2} \\
B_n(v,w;0)&=& \frac{v^{n+1}-w^{n+1}}{v-w}
\end{eqnarray}
where now $v=1/\beta-1$ and $w=1/\alpha-1$. It can be verified, using
the identity
\begin{equation}
\sum_{n=r}^{X-Y} (-1)^n \binom{X}{Y-n} \binom{n}{r} = (-1)^r
\binom{X-1-r}{Y-r} \;,
\end{equation}
that (\ref{eqn:Zsumrep}) can be rewritten as 
\begin{equation}
Z_N = \braket{W}{V} \sum_{k=0}^{N}   
\left[ \binom{2N-2-k}{N-k} - \binom{2N-2-k}{N-2-k} \right]
\ \left[ \frac{(1/\beta)^{k+1} - (1/\alpha)^{k+1}}{1/\beta-1/\alpha} \right]
\end{equation}
which is equivalent to equation (39) of \cite{DEHP}.

We note that for $q \to 1$ the singularity in the denominator of
(\ref{eqn:Zsumrep}) is cancelled by the sum over
$R_{N,n}(q)B_n(v,w;q)$ and the expression is in fact well-behaved.
Although we have checked for small system sizes that
(\ref{eqn:Zsumrep}) agrees with the expression of \cite{SMW} we have
not been able to show this in a simple way.

\section{The phase diagram of the model}
\label{sec:phasediag}

We now have enough information to obtain an exact phase diagram for
the model and expressions for the particle current in the large system
size limit. The behaviour differs greatly according to whether the
particles are forward biased ($q<1$) or reverse biased ($q>1$) and so
we treat the two cases separately.

\subsection{The forward bias regime}
\label{sec:forbias}

When $q<1$, the quantities of interest are most quickly obtained from
the integral (\ref{eqn:Zint:q<1:long}) as was also done in
\cite{Sasamoto1}. As $N$ becomes large, we can use the saddle-point
method to evaluate $Z_N$, and so we re-write (\ref{eqn:Zint:q<1:long})
as a contour integral
\begin{equation}
\label{eqn:Zint:q<1:complex}
Z_N = \frac{(q;q)_\infty}{4\pi i} \left( \frac{1}{1-q} \right)^{\!\!N}
\oint_K \frac{\D z}{z} (2+z+z^{-1})^N \,
\frac{(z^2,z^{-2};q)_\infty}
{(vz,wz,vz^{-1},wz^{-1};q)_\infty}
\end{equation}
where the contour $K$ is the circle $|z|=1$ and is directed anti-clockwise.
Furthermore it passes through the saddle-point of $2+z+1/z$ along the
path of steepest descent. We find from the saddle-point formula that
\begin{equation}
\label{eqn:Z_N:maxcurrent}
Z_N \sim \frac{4}{\sqrt{\pi}} \frac{(q;q)_\infty^3}{(v,w;q)_\infty^2}
\left( \frac{1}{N} \right)^{\!\!\frac{3}{2}} \left( \frac{4}{1-q}
\right)^{\!\!N}
\end{equation}
which holds as long as $|v|<1$ and $|w|<1$.

\begin{figure}[htb]
\begin{center}
\includegraphics[scale=1]{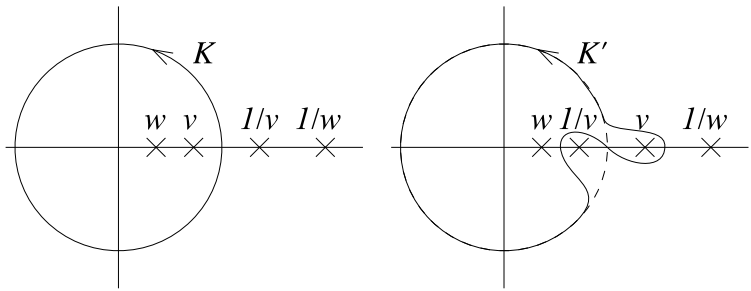}
\end{center}
\caption{\label{fig:contour}The integral over the contour $K$ (shown
  dotted in the right-hand figure) can be determined from a
  saddle-point expansion, whereas the distorted contour $K^\prime$ is
  the correct one to use when $1<v<1/q$. Note that only the four poles
  closest to $z=1$ have been shown for clarity.}
\end{figure}

We can treat other values of $v$ and $w$ using the integral
(\ref{eqn:Zint:q<1:complex}) after realising that
(\ref{eqn:Z_N:maxcurrent}) gives the contribution from the contour $K$
for any value of $v$ and $w$, whereas the analytic continuation of
$Z_N$ is obtained by distorting the contour such that the poles at
$z=v,qv,q^2v,\ldots$ and $z=w,qw,q^2w,\ldots$ stay inside it and all
other poles are outside it. Figure \ref{fig:contour} illustrates how
to modify the contour as $v$ is increased to a value $1<v<1/q$.

The difference between the two results can be calculated using the
residue theorem. When $|w|<1$, $1<v<1/q$ we find by looking at the poles
at $z=v$ and $z=1/v$ that we should add
\begin{equation}
\label{eqn:dominant}
\frac{(v^{-2};q)_\infty}{(vw,w/v;q)_\infty}
\left(\frac{2+v+v^{-1}}{1-q}\right)^{\!\!N}
\end{equation}
to (\ref{eqn:Z_N:maxcurrent}). In the large $N$ limit, this correction
dominates the contribution from the contour $K$ and so we write
\begin{equation}
\label{eqn:Z_N:hidensity}
Z_N \sim 
\frac{(v^{-2};q)_\infty}{(vw,w/v;q)_\infty}
\left(\frac{2+v+v^{-1}}{1-q}\right)^{\!\!N}
\end{equation}
when $|w|<1$ and $1<v<1/q$. As $v$ is increased above $1/q$ and other
poles need to be considered, (\ref{eqn:dominant}) remains the dominant
contribution to $Z_N$. One could guess this from the fact that the
pole at $z=v$ is furthest from the origin.

Due to the symmetry of (\ref{eqn:Zint:q<1:complex}) in $v$ and $w$, we
obtain (\ref{eqn:Z_N:hidensity}) with $v \leftrightarrow w$ when
$w>1$, $|v|<1$. When both $w$ and $v$ are greater than one, the leading
term in the asymptotic expansion comes, as before, from the pole
furthest along the real axis of the complex plane. Thus we have found
three different forms for $Z_N$, each of which corresponds to a phase
in the model. These forms and their regions of validity are presented
in table \ref{table:Z_N:q<1}.

\begin{table}[htb]
\begin{center}
\begin{tabular}{c|c}
Region & Normalisation $Z_N$ \\[0.5ex] \hline 
& \\
$v<1$, $w<1$ & 
$\displaystyle
\frac{4}{\sqrt\pi}\frac{(q;q)_\infty^3}{(v,w;q)_\infty^2} 
\left( \frac{1}{N} \right)^{\!\!\frac{3}{2}}
\left( \frac{4}{1-q} \right)^{\!\!N}$
\\[3ex]
$v>w$, $v>1$ & 
$\displaystyle \frac{ (v^{-2};q)_\infty }{ (vw,w/v;q)_\infty }
\left( \frac{2+v+v^{-1}}{1-q} \right)^{\!\!N}$
\\[3ex]
$w>v$, $w>1$ &
$\displaystyle \frac{ (w^{-2};q)_\infty }{ (wv,v/w;q)_\infty }
\left( \frac{2+w+w^{-1}}{1-q} \right)^{\!\!N}$ 
\\[3ex]
\end{tabular}
\end{center}
\caption{\label{table:Z_N:q<1}The normalisation for different values
  of $v$ and $w$ when $q<1$}
\end{table}

We can go on to find the currents in the three forward bias phases
through equation (\ref{eqn:jdef}). These expressions are presented in
table \ref{table:j:q<1}. For completeness we should determine $Z_N$
along each of the phase boundaries ($v=w>1$ or $v=1, w\ne1$ etc). We
find that the currents subsequently found are equal to the limiting
values of those in table \ref{table:j:q<1} as the boundary under
consideration is approached from each of the neighbouring regions.

\begin{table}[htb]
\begin{center}
\begin{tabular}{c|c}
Region & Current $J$\\[0.5ex] \hline 
& \\
$\alpha>\frac{1-q}{2}$, $\beta>\frac{1-q}{2}$ &
 $\displaystyle \frac{1-q}{4}$\\[3ex]
$\alpha<\frac{1-q}{2}$, $\beta>\alpha$ &
$\displaystyle \frac{\alpha(1-q-\alpha)}{1-q}$\\[3ex]
$\beta<\frac{1-q}{2}$, $\alpha>\beta$ &
$\displaystyle \frac{\beta(1-q-\beta)}{1-q}$\\[3ex]
\end{tabular}
\end{center}
\caption{\label{table:j:q<1}The $N\to\infty$ forms of the particle
  current in the forward biased phases.}
\end{table}

\begin{figure}[htb]
\begin{center}
\includegraphics[scale=1.5]{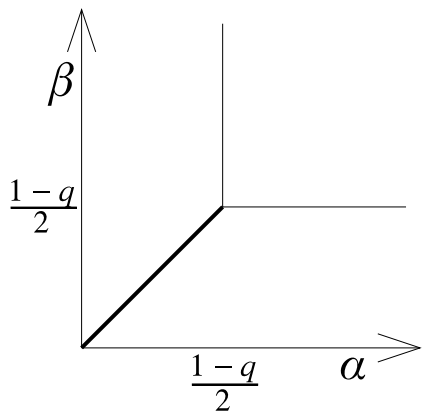}
\end{center}
\caption{\label{fig:phasediag:q<1}The phase diagram of the model when
  $q<1$. The thick solid line is a first order transition and the thin
  solid lines are second order transitions in the sense of \cite{DEHP,SD}.}
\end{figure}

We can now draw a phase diagram for the system when $q<1$: see figure
\ref{fig:phasediag:q<1}. We note that it has the same structure as
that found for $q=0$ \cite{DEHP}. The region $|v|<1$, $|w|<1$ is a
maximal current phase; the remaining two phases correspond to the high
and low density phases found by \cite{DEHP}. Each of these two latter
phases may be subdivided into three regions according to the
behaviour of the density correlation length in the thermodynamic limit
\cite{Sasamoto1}.

\subsection{The reverse bias regime}
\label{sec:revbias}

We turn now to the case $q>1$. We will show shortly by examining the
exact formula (\ref{eqn:Zsumrep}) that the normalisation behaves like
$Z_N \sim q^{\frac{1}{4}N^2}$ for large $N$. This sheds further light
on why the saddle-point method is not applicable here: by its nature,
it gives expressions where the exponent is linear in $N$ rather than
the desired quadratic.

To proceed we must find approximate forms of two elementary
quantities: the $q$-shifted factorial and the $q$-binomial
coefficient. We rewrite the definition (\ref{eqn:qfacdef}) as
\begin{equation}
\label{eqn:qfac:approx}
(q;q)_n = (-1)^n q^{\binom{n+1}{2}} e^{M_n(q)}
\end{equation}
where
\begin{equation}
M_n(q)=-\sum_{k=1}^\infty \frac{1}{k} \frac{1}{q^k-1} (1-q^{-kn}) \;.
\end{equation}
We see that when $n$ is large, $M_n(q)$ can be approximated by
\begin{equation}
\label{eqn:M:approx}
M(q) \simeq -\sum_{k=1}^\infty \frac{1}{k} \frac{1}{q^k-1}
\end{equation}
which is independent of $n$. This leads to an approximation of the
$q$-binomial coefficient
\begin{equation}
\label{eqn:qbinomial:approx} 
\qbinom{n}{k} \simeq q^{k(n-k)} e^{-M(q)}
\end{equation}
which is valid when both $n$ and $k$ are large. The rest of the
analysis is not particularly illuminating, so is presented in appendix
C. We ultimately find that when $q>1$
\begin{equation}
\label{eqn:Z:q>1}
Z_N \sim A(v,w;q) \, (\invq vw,1/vw;\invq)_\infty
     \left( \frac{\sqrt{vw}}{q-1} \right)^{\!\!N} q^{\frac{1}{4}N^2}
\end{equation}
where
\begin{equation}
\label{eqn:Adef}
A(v,w;q) = \sqrt{\frac{\pi}{\ln q}} \exp \left\{ M(q)+\frac{(\ln
    w/v)^2}{4 \ln q} \right\} \;.
\end{equation}
We should note that the smallest system size $N$ for which
(\ref{eqn:Z:q>1}) holds will be a function of $q$ due to difference
between the exact quantity $M_n(q)$ and the approximate form we used
$M(q)$.

The expression for the current follows quickly from (\ref{eqn:jdef}).
This reads
\begin{equation}
\label{eqn:j:q>1}
J \sim \left( \frac{\alpha\beta(q-1)^2}{(q-1+\alpha)(q-1+\beta)}
\right)^{\!\!\frac{1}{2}} q^{-\frac{1}{2}N+\frac{1}{4}}
\end{equation}
which, in contrast with the currents in the forward bias regime, is a
function of the number of lattice sites $N$.

\section{Discussion}
\label{sec:conclude}

In this work we have employed properties of $q$-Hermite polynomials to
calculate exact steady state properties of the partially asymmetric
exclusion process.  The connection between this model and $q$-Hermite
polynomials lies in the fact that the matrices $D$ and $E$ of the
matrix product solution can be written in terms of $q$-raising and
$q$-lowering operators as in (\ref{eqn:Dofa}).  Then the calculation
of the normalisation (\ref{eqn:Zdef}) amounts to decomposing the
vectors $\langle W |$ and $|V \rangle$ onto the eigenvectors of the
matrix $C$ which are the eigenvectors of the ``co-ordinate'' operator
of the q-deformed oscillator.  This allowed us to obtain integral
representations of the normalisation for both the forward bias case
$q<1$ (\ref{eqn:Zint:q<1:short}) and the reverse bias case $q>1$
(\ref{eqn:Zint:q>1:short}).  Further we could use orthogonality
properties of the $q$-Hermite polynomials to express these two
integral expressions as a finite sum valid for all $q$
(\ref{eqn:Zsumrep}).

In a very recent paper \cite{Sasamoto1}, orthogonal polynomials were
also used to study the ASEP. In view of the fact that our work and
\cite{Sasamoto1} were carried out independently, a comparison is in
order.  In \cite{Sasamoto1}, the large system size limits of the
normalisation and the current in the forward bias regime $q<1$ via the
integral representation (\ref{eqn:Zint:q<1:short}) were obtained.
Furthermore it was shown that one could also analyse the density
correlations from this integral. In the present work we have found
that a corresponding integral can also be found for the case of
reverse bias $q>1$. Further for all values of $q$ (and $\alpha$,
$\beta$) we have succeeded in obtaining an exact sum formula valid for
all system sizes. One application of this general expression was to
calculate the current in the reverse bias phase.

For the forward bias case the phase diagram proposed by Sandow
\cite{Sandow} is recovered. In that work the more general parameter
space including rates $\gamma$ (exit of particles at the left
boundary) and $\delta$ (entry at the right) was considered. For this
case the algebra is modified \cite{DEHP} to
\begin{eqnarray}
DE-qED &=& D+E \\
\bra{W}(\alpha E - \gamma D) &=& \bra{W} \\
(\beta D - \delta E) \ket{V} &=& \ket{V} \;.
\end{eqnarray}
In principle we can generalise our method to that case. The only
difference being that $\langle W |$ and $|V \rangle$ specified by
(\ref{eqn:EonW}) and (\ref{eqn:DonV}) will have more complicated
expressions than (\ref{eqn:Vofn}) and (\ref{eqn:Wofn}). In the forward
bias case the generalisation would not produce any new phases.
However for $q>1$ allowing particles to exit at the left and enter at
the right (both $\gamma, \delta > 0$) would allow a left flowing current
of particles to be sustained, thus destroying the reverse bias phase.

The reverse bias phase where the boundary conditions impose a current
opposite to the bulk bias realises a new phase in the ASEP where the
current decreases exponentially with system size. A typical
arrangement of particles in this phase is a lattice full at the left
end and empty at the right end.  The form for the current $j \sim
q^{-N/2}$ (\ref{eqn:j:q>1}) suggests that the lattice is typically
half-full {\it i.e.}  the furthest particle to the right has typically
to traverse a distance of $N/2$ sites against the bias to exit the
lattice.  To understand why the lattice is half full one invokes the
particle hole symmetry that the current of particles exiting to the
right must equal the current of holes exiting to the left.  The
symmetry implies that the lattice must be half-full \cite{RB}.

Further, for large $N$ the current tends to zero and one can compare
with the much simpler case where the current is exactly zero, for
example when the boundaries are reflecting \cite{SchutzSandow}. Then
the microscopic dynamics obey detailed balance and the unnormalised
probability of a configuration of $M$ particles at positions $x_1,
x_2, \ldots, x_M$ will be proportional to $q^{-\sum_{i=1}^{M} x_i}$.
Here also the normalisation grows exponentially in $N^2$ \cite{EKKM}.

In a recent preprint \cite{Sasamoto2} the density profiles for the
three forward bias $q<1$ phases were calculated in the thermodynamic
limit. In particular in the maximal current phase it was shown that
the density profile decays with distance $x$ from the left boundary as
$1/2+(4\pi x)^{-1/2}$ for large $x$.  There remain, however, a number
of issues to be resolved. As $q$ tends to $1$ the maximal current
phase occupies more and more of the phase diagram---see
figure~\ref{fig:phasediag:q<1}. However, at $q=1$ we know that the
profile is exactly linear. This implies a non-trivial limit $q \to 1$
and therefore non-trivial crossover phenomena from the asymmetric to
the symmetric case. This corresponds to the transition between KPZ and
EW universality classes in the related growth models.  Further in the
reverse bias case ($q>1$), as we expect the lattice to be roughly
half-full, the density profile should be sigmoid-like. For $q \to
\infty$ the profile will be a sharp step function whereas as $q \to 1$
the sigmoid profile will straighten out into a linear profile.

It is known that the quadratic algebra
(\ref{eqn:DEcommute}--\ref{eqn:DonV}) for the open boundary problem
can be used to solve a partially asymmetric periodic system with the
addition of defect particles \cite{DJLS,DerridaReview}. A defect
particle hops forward with rate $\alpha$ but is overtaken (and moved
back a site) by normal particles with rate $\beta$. When $q<1$, the
different phases in the present problem manifest themselves in this
model \cite{Sasamoto3}. However the case $q>1$ is yet to be tackled;
we believe this to be of special interest as the reverse bias phase
corresponds to phase separation into pure domains and spontaneous
breaking of translational invariance \cite{EKKM,AHR}.

\section*{Acknowledgments}

Our warm thanks go to Magnus Richardson who contributed to the early
development of this work. Also we thank Mike Gunn who helped us with
equation (\ref{eqn:qfac:approx}). RAB acknowledges an EPSRC
studentship and MRE thanks the Royal Society for a University Research
Fellowship. FC thanks the EPSRC for financial support and FHLE is an
EPSRC Advanced Fellow.


\section*{Appendix A: Generating functions of the $q$-Hermite polynomials}
\label{appx:genfunc}
\setcounter{equation}{0}
\def\theequation{A\arabic{equation}}

In this appendix we explain how to obtain the generating functions
$G(\theta,\lambda)$ and $G(u,\lambda)$ from the recursion relations
for the $q$-Hermite polynomials (\ref{eqn:hprecrel}). We also present
explicit expressions for the polynomials as they may be obtained
easily from the generating functions.

We consider first a general form of $G$, suitable for both $q<1$ and
$q>1$:
\begin{equation}
\label{eqn:Gdef}
G(x,\lambda)=\sum_{n=0}^{\infty} \frac{\lambda^n}{\sqrt{(q;q)_n}}
\braket{x}{n} \;.
\end{equation}
We now obtain a functional relation for $G(x,\lambda)$ by multiplying
both sides of equation (\ref{eqn:hprecrel}) by
$\lambda^n/(\sqrt{(q;q)_n})$ and performing the required summations:
\begin{equation}
\label{eqn:Gfuncrel}
G(x,q\lambda)=(\lambda^2 - 2\lambda x + 1) G(x,\lambda) \;.
\end{equation}
By using this relation repeatedly, we can find an expression for
$G(x,\lambda)$ in terms of $G(x,0)$. This latter quantity is fixed by
normalisation, and so we set it to $1$.

It can be seen from (\ref{eqn:Gfuncrel}) that our approach from
$G(x,\lambda)$ to $G(x,0)$ depends on whether $q<1$ or $q>1$. Consider
first the case $q<1$. It is useful to make the change of variable
$x=\cos\theta$ so that (\ref{eqn:Gfuncrel}) becomes
\begin{equation}
  G(\theta,\lambda)=\frac{G(\theta,q\lambda)}{(1-\lambda
    e^{i\theta})(1-\lambda e^{-i\theta})} \;.
\end{equation}
Iterating this we find
\begin{equation}
\label{eqn:Gnice}
G(\theta,\lambda)=\frac{1}{(\lambda e^{i\theta},\lambda
  e^{-i\theta};q)_\infty}
\end{equation}
where we have used $G(\theta,0)=1$. The infinite product
$1/(x;q)_\infty$ has a well-known series representation
\cite{GaspRah} valid for $x<1$, $q<1$
\begin{equation}
\frac{1}{(x;q)_\infty}=\sum_{n=0}^{\infty} \frac{x^n}{(q;q)_n}
\end{equation}
from which we may extract the form of $\braket{\theta}{n}$. Expanding
both sides of (\ref{eqn:Gnice}) in $\lambda$ and comparing co-efficients
we find
\begin{equation}
\label{eqn:HP:q<1}
\braket{\theta}{n}=\frac{1}{\sqrt{(q;q)_n}} \sum_{k=0}^{n}
\qbinom{n}{k} e^{i(n-2k)\theta}
\end{equation}
where $\qbinom{n}{k}$ is the $q$-deformed binomial described in
section \ref{sec:sumformula}.

The case $q>1$ proceeds in the same way. We must however make a
different change of variable $x= i\sinh u$ because otherwise
(\ref{eqn:xonx}) would imply that we had found imaginary eigenvalues
of a Hermitian matrix. Also we should divide $\lambda$ by $q$ in as we
approach $G(u,0)$ from $G(u,\lambda)$. We thus re-write
(\ref{eqn:Gfuncrel}) as
\begin{equation}
  G(u,\lambda)=(1-i\invq\lambda e^u) (1+i\invq\lambda e^{-u})
  G(u,\invq\lambda)
\end{equation}
and iterate before to obtain
\begin{equation}
\label{eqn:Gnicer}
G(u,\lambda)=(i\invq\lambda e^u,-i\invq\lambda e^{-u};\invq)_\infty \;.
\end{equation}
Again the infinite product on the right-hand side of this equation has
a useful series expansion appropriate for $q>1$ and all $x$:
\begin{equation}
(\invq x;\invq)_\infty = \sum_{n=0}^{\infty} \frac{x^n}{(q;q)_n} \;.
\end{equation}
Expansion of \ref{eqn:Gnicer} in powers of $\lambda$ and comparison
with the generating function (\ref{eqn:genfunc:q>1}) yields
$\braket{u}{n}$:
\begin{equation}
\label{eqn:HP:q>1}
\braket{u}{n}=\frac{i^n}{\sqrt{(q;q)_n}} \sum_{k=0}^{n}
(-1)^k \qbinom{n}{k} e^{(n-2k)u} \;.
\end{equation}

It is important to realise that the two forms (\ref{eqn:HP:q<1}) and
(\ref{eqn:HP:q>1}) we have found are not very different. In particular
one can obtain the form (\ref{eqn:HP:q>1}) by making the substitution
$\theta \to \pi/2-iu$ which is another way of describing the
replacement $\cos\theta \to i\sinh u$. Also we should note that all
the functions we have found are real on their domains despite the
presence of $i$ when $q>1$.

\section*{Appendix B: Expansion of the cosine function in $q$-Hermite polynomials}
\label{appx:cosexpand}
\setcounter{equation}{0}
\def\theequation{B\arabic{equation}}

Here we show how to re-write $[2(1+\cos\theta)]^N$ as a sum of
$q$-Hermite polynomials $\braket{\theta}{n}$. First of all we simplify
the task by using an identity easily verified by induction:
\begin{equation}
\label{eqn:cossimplify}
[2(1+\cos\theta)]^N = \sum_{n=0}^{N} \binom{2N}{N-n} c_n(\theta)
\end{equation}
with
\begin{equation}
c_n(\theta)=
\left\{ \begin{array}{ll}
  1 & n=0 \\
  2\cos(n\theta) & n>0 \;.
\end{array} \right.
\end{equation}

We now need only to consider the expansion of $c_n(\theta)$. It is
fairly easy to convince oneself by inspecting $(\ref{eqn:HP:q<1})$
that only those $q$-Hermite polynomials of the same parity as the
cosine function $c_n$ will appear in the expansion. Also we do not
expect any contributions from $\braket{\theta}{k}$ with $k>n$. This
leads to the following prescription:
\begin{equation}
\label{eqn:cexpand}
c_n(\theta)=\sum_{k=0}^{\halfint{n}} a_{n,k}
\braket{\theta}{n-2k} \;.
\end{equation}

A formula for $a_{n,k}$ may be found by applying the orthogonality
property of the $q$-Hermite polynomials (\ref{eqn:orthog:q<1}). We
obtain a familiar integral transform
\begin{equation}
\label{eqn:inttrafo}
a_{n,k}= \int_{0}^{\pi} \Di{\theta} \nu(\theta) c_n(\theta)
\braket{n-2k}{\theta} \;.
\end{equation}
To evaluate this integral we need the series expansion of the weight
function $\nu(\theta)$
\begin{equation}
\nu(\theta)=\frac{1}{2\pi} \sum_{s=-\infty}^{\infty} (-)^s
q^{\binom{s}{2}} \left( 1+q^s \right) e^{2is\theta} =
\frac{1}{2\pi} \sum_{s=-\infty}^{\infty} b_s e^{2is\theta} \;.
\end{equation}
Inserting this and the explicit formula for the $q$-Hermite polynomial
(\ref{eqn:HP:q<1}) into the above we find, after some manipulation,
\begin{equation}
a_{n,k} = \frac{1}{\sqrt{(q;q)_{n-2k}}} \sum_{s=-\infty}^{\infty}
b_s \sum_{r=0}^{n-2k} \qbinom{n-2k}{r}
\frac{1}{2\pi} \int_{0}^{2\pi} \Di{\theta} \cos((k+r-s)\theta) \;.
\end{equation}

The integral that appears in this equation is just a representation of
the Kronecker delta symbol $\delta_{s,k+r}$. Thus we can eliminate the
summation over $s$:
\begin{eqnarray}
a_{n,k} &=& \frac{(-1)^k}{\sqrt{(q;q)_{n-2k}}} \sum_{r=0}^{n-2k}
\qbinom{n-2k}{r} (-1)^r q^{\binom{k+r}{2}} (1+q^{k+r}) \nonumber\\
&=& \frac{(-1)^k q^{\binom{k}{2}}}{\sqrt{(q;q)_{n-2k}}} \sum_{r=0}^{n-2k}
\qbinom{n-2k}{r} (-1)^r q^{\binom{r}{2}+kr} \left(1+q^{k+r} \right)
\nonumber\\
&=& \frac{(-1)^k q^{\binom{k}{2}}}{\sqrt{(q;q)_{n-2k}}}
\left( (q^k;q)_{n-2k} + q^k (q^{k+1};q)_{n-2k} \right) \;.
\end{eqnarray}
To do the last step, we have made used the following series expansion
in $x$:
\begin{equation}
(x;q)_n = \sum_{k=0}^{n} \qbinom{n}{k} (-1)^k q^{\binom{k}{2}} x^k \;.
\end{equation}

The latest expression allows a little simplification by noting that
\begin{equation}
(aq^k;q)_{n-2k} = \frac{(a;q)_{n-k}}{(a;q)_k}
\end{equation}
and so we find we find
\begin{equation}
\label{eqn:afinal}
a_{n,k} = (-1)^k \sqrt{(q;q)_{n-2k}} \, q^{\binom{k}{2}} \left(
\qbinom{n-k-1}{k-1} + q^k \qbinom{n-k}{k} \right)
\end{equation}
which is true for all $n$, $k$ if we observe the usual convention that
$\qbinom{n}{k}=0$ when $k<0$ or $k>n$.

We may now combine equations (\ref{eqn:cossimplify}),
(\ref{eqn:cexpand}) and (\ref{eqn:afinal}) to find
\begin{eqnarray}
[2(1+\cos\theta)]^N &=& \sum_{n=0}^{N} \binom{2N}{N-n}
\sum_{k=0}^{\halfint{n}} a_{n,k}(q) \braket{\theta}{n-2k} \nonumber\\
&=& \sum_{n=0}^{N} \sum_{k=0}^{\halfint{N-n}}
\binom{2N}{N-(n+2k)} a_{n+2k,k} \braket{\theta}{n} \nonumber\\
&=& \sum_{n=0}^{N} R_{N,n}(q) \sqrt{(q;q)_n} \, \braket{\theta}{n}
\end{eqnarray}
thus completing the derivation of equation (\ref{eqn:cosexpand}). The
problem of expanding of a general function in $q$-Hermite polynomials
was first solved by Rogers in 1894. His approach, detailed by
\cite{Andrews}, is more complicated than ours as he did not use the
orthogonality properties of the $q$-Hermite polynomials.

\section*{Appendix C: Approximation of the normalisation for large $N$ and $q>1$}
\label{appx:q>1}
\setcounter{equation}{0}
\def\theequation{C\arabic{equation}}

We indicate here how to estimate $Z_N$ as given by (\ref{eqn:Zsumrep}) in a
systematic manner when $N$ is large and $q>1$. We begin with the function
$B_n(v,w;q)$ which is defined by the sum in equation (\ref{eqn:Bdef}). The
dominant terms are those around $k=n/2$, and so we may replace the
$q$-binomial with the approximation (\ref{eqn:qbinomial:approx}) and also
rewrite the sum as an integral over $k$. This gives for large $n$
\begin{equation}
\label{eqn:Bapprox}
B_n(v,w;q) \sim (-1)^n A(v,w;q)\, q^{\frac{1}{4}n^2} |vw|^{\frac{1}{2}n}
\end{equation}
where $A(v,w;q)$ is given by (\ref{eqn:Adef}).

We now consider the sum (\ref{eqn:Rdef}) for $R_{N,n}$. In this summation
we keep only the term with largest $k$ as the others are exponentially
suppressed. We find then that
\begin{equation}
Z_N \simeq \frac{(\invq vw;\invq)_\infty}{(1-q)^N} \left( S_N(v,w;q)
  + 2N S_{N-1}(v,w;q) \right)
\end{equation}
where we have expanded the product $\braket{W}{V}$ using the identity
\begin{equation}
\label{eqn:exp:q>1}
\sum_{r=0}^\infty \frac{(qx)^r}{(q;q)_r} = (x;\invq)_\infty
\end{equation}
which holds when $q>1$ and for all $x$ and where we have defined
\begin{equation}
S_N(v,w;q) = (1-q^N) \sum_{r=0}^{\halfint{N}} (-1)^r
q^{\binom{r}{2}} \frac{(q;q)_{N-r-1}}{(q;q)_r (q;q)_{N-2r}}
B_{N-2r}(v,w;q)
\;.
\end{equation}

The main contribution to this latest sum $S_N(v,w;q)$ is where $r$ is small.
The approximation
\begin{equation}
(-1)^r q^{\binom{r}{2}} \frac{(q;q)_{N-r-1}}{(q;q)_r (q;q)_{N-2r}} \simeq
- \frac{q^{-r^2+(N+1)r-N}}{(q;q)_r}
\end{equation}
which follows from (\ref{eqn:qfac:approx}) and (\ref{eqn:M:approx}) is
valid in that region and when combined with the asymptotic expression
for $B_n(v,w;q)$ yields
\begin{equation}
S_N(v,w;q) \sim (-1)^N A(v,w;q) \, (1-q^{-N}) |vw|^{\frac{1}{2}N}
q^{\frac{1}{4}N^2} \sum_{r=0}^{\halfint{N}} \frac{1}{(q;q)_r} \left(
  \frac{q}{vw} \right)^{\!\!r}
\;.
\end{equation}
We are now left with a single summation which may be estimated from
the identity (\ref{eqn:exp:q>1}). We see
\begin{equation}
\sum_{r=0}^{\halfint{N}} \frac{1}{(q;q)_r} \left(
  \frac{q}{vw} \right)^{\!\!r} = (1/vw;\invq)_\infty +
  \mathcal{O}(q^{-\frac{1}{4}N^2})
\end{equation}
and so to leading order in $q$ we find
\begin{equation}
S_N(v,w;q) \sim (-1)^N A(v,w;q) \, (1/vw;\invq)_\infty |vw|^{\frac{1}{2}N}
q^{\frac{1}{4}N^2}
\end{equation}

Noting that $S_{N-1}$ is exponentially smaller than $S_N$ we may finally
write down the asymptotic form of $Z_N$ when $q>1$
\begin{equation}
Z_N \sim A(v,w;q) \, (\invq vw,1/vw;\invq)_\infty
     \left( \frac{\sqrt{vw}}{q-1} \right)^{\!\!N} q^{\frac{1}{4}N^2}
\end{equation}
which is the expression presented in section \ref{sec:revbias}.


\end{document}